\documentclass[aps,prapplied,reprint,groupedaddress,showpacs]{revtex4-2}
\usepackage{graphicx,graphics}
\usepackage[table]{xcolor}
\usepackage{epstopdf}
\usepackage{amsmath,amssymb,amsfonts}
\usepackage{latexsym,verbatim}
\usepackage{natbib}
\usepackage{bm}
\usepackage{bbold}
\usepackage{color}
\usepackage{ulem}
\usepackage[breaklinks=false,colorlinks,citecolor=blue,linkcolor=blue,urlcolor=blue]{hyperref}
\usepackage[abs]{overpic}
\usepackage{textcomp} 
\usepackage{mathtools}
\usepackage{braket}

\begin{document}

\title{Towards hole-spin qubits in Si pMOSFETs within a planar CMOS foundry technology}
\author{L. Bellentani$^1$, M. Bina$^2$, S. Bonen$^3$, A. Secchi$^1$, A. Bertoni$^1$, S. Voinigescu$^3$, A. Padovani$^2$, L. Larcher$^2$, and F. Troiani$^1$}
\affiliation{$^1$S3, Istituto Nanoscienze-CNR, Modena, Italy}
\affiliation{$^2$Applied Materials - MDLx Italy R\&D, Reggio Emilia, Italy}
\affiliation{$^3$Edward S. Rogers Snr. Department of Electrical and Computer Engineering, University of Toronto, Toronto, Canada}
\begin{abstract}

Hole spins in semiconductor quantum dots represent a viable route for the implementation of electrically controlled qubits. 
In particular, the qubit implementation based on Si pMOSFETs offers great potentialities in terms of integration with the control electronics and long-term scalability. 
Moreover, the future down scaling of these devices will possibly improve the performance of both the classical (control) and quantum components of such monolithically integrated circuits. 
Here we use a multi-scale approach to simulate a hole-spin qubit 
in a down scaled Si-channel pMOSFET, whose structure is based on a commercial 22nm fully-depleted silicon-on-insulator device. 
Our calculations show the formation of well defined hole quantum dots within the Si channel, and the possibility of a general electrical 
control, with Rabi frequencies of the order of $100\,$MHz for realistic field values. Our calculations demonstrate the crucial role of the channel aspect ratio, and the presence of a favorable parameter range for the qubit manipulation. 

\end{abstract}

\date{\today}

\maketitle

\section{Introduction}


Localized spins in semiconductors were early recognized as one of the most promising means for the encoding and manipulation of quantum information \cite{Loss1998a}. 
In the last decade, this approach has gained a renewed interest, thanks to the high degree of control achieved on the single- and few-particle states in Si \cite{Zwanenburg2013a} and Ge quantum dots \cite{Scappucci2020a}. 
Within this platform, all the fundamental criteria for the implementation of quantum computation have been recently met, from  high-fidelity one- \cite{Veldhorst2014a,Kawakami2016a,Yoneda2018a} and two-qubit gates \cite{Zajac2018a,Watson2018a,Huang2019a}, to qubit read out \cite{Zheng2019a,Keith2019a,Urdampilleta2019a} and coherent spin transfer \cite{Kandel2019a,Mills2019a}. 
Crucially, carrier spins in Si and Ge benefit from long coherence times \cite{Muhonen2014a}, thanks to the limited impact of hyperfine 
interactions, which represent instead an ubiquitous source of decoherence in III-V compounds. 
Besides, Si and Ge represent key 
materials in modern electronics, and thus provide a common platform for integrating qubits and control circuits \cite{Veldhorst2017a,Pauka2019a,Xue2020a}.

\begin{figure}[t]
\includegraphics[width=1.\columnwidth]{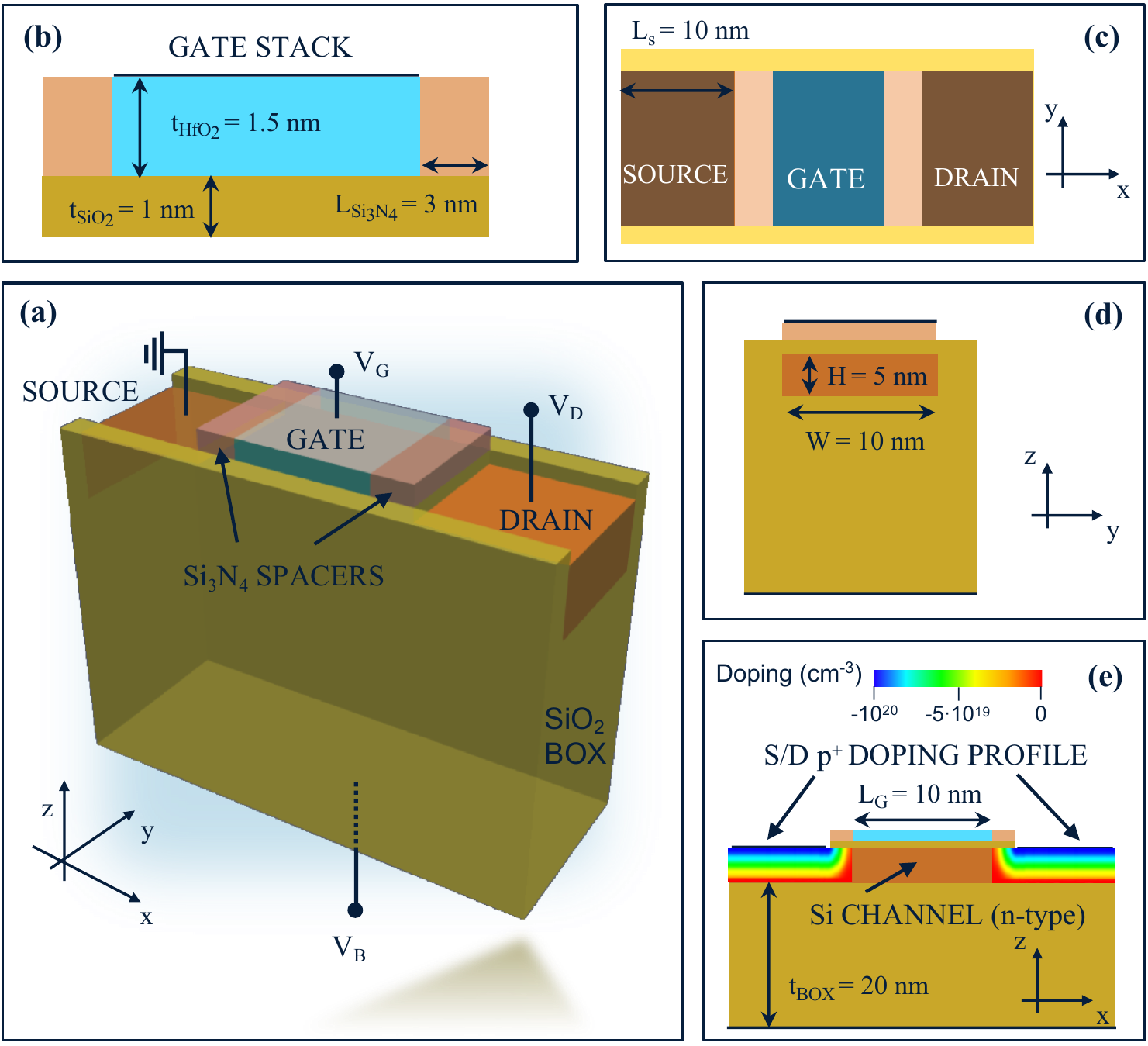}
\caption{(a) Schematics of the simulated pMOSFET FDSOI transistor. (b) Composition and geometry of the high-$\kappa$ metal gate stack. Top (c), side (d), and front (e) views of the device; in the latter we highlight the non-uniform Gaussian doping profile in the source and drain.}
\label{fig:device}
\end{figure}
As regards scalability, a fundamental advantage of the semiconductor-based approach would be represented by the use of homogeneous building blocks for realizing both the quantum (qubits) and classical (control electronics) components of the hardware. This route has been undertaken with the implementation of complementary metal–oxide–semiconductor (CMOS) qubits, obtained by suitably modifying transistors currently fabricated in foundry processes \cite{Maurand2016a,Veldhorst2017a,Crippa2018a,Bonen2019a,Urdampilleta2019a}.   
In particular, hole spins in group-IV materials offer strong spin-orbit coupling, which enables the all-electric qubit manipulation, thus avoiding the introduction in the circuit design of elements that are incompatible with the industrial fabrication processes
\cite{Kloeffel2013a,Maurand2016a,Watzinger2018a,Crippa2019a,Hendrickx2020a,Hendrickx2020b,Scappucci2020a}.

One of the key challenges in this direction is to find a trade-off amongst the conflicting requirements of operating the qubits at milli-Kelvin temperatures, in order to enhance the coherence times, and the control electronics at higher temperatures (at least $1-4\,$K), in order to allow a sufficiently fast removal of the dissipated power \cite{Vandersypen2017a,Ono2019a,West2019a,Petit2020a,Yang2020a}. In this respect, a crucial role might be played by the qubit implementation in down scaled MOSFETs, which would improve the efficiency of the control and read-out electronics, and enhance, with the quantum confinement, the temperatures and frequencies at which the qubits can be operated.

In this paper, we perform a multi-scale simulation of fully depleted silicon on insulator pMOSFETs, which represents a down scaled version of $22\,$nm commercial devices. The TCAD simulation of the devices allows us to account for the material properties, doping profile, and electrical parameters in the device, and to derive realistic profiles for the confining  potential in the low-temperature regime. The hole states are then computed within a 6 bands $\bf{k}\cdot\bf{p}$ approach, in order to derive the quantities that define the properties and functionalities of the spin qubit. Our simulations show that the considered device geometry allows the formation of a well defined quantum dot, entirely localized at the center of the Si channel.  Besides, the application to the top gate of different voltage pulses enables the implementation of spin-qubit rotations around all three axes of the Bloch sphere, with Rabi frequencies of the order of $10^2\,$MHz. We determine the dependence of such Rabi frequencies on the orientation of the static magnetic field, and find a trade-off between optimizing longitudinal and transverse rotations. Finally, we identify the optimal aspect ratio of the Si channel, which corresponds to the occurrence of a clear transition in the ground state properties.

The paper is organized as follows. Section II is devoted to the modeling of the scaled pMOSFET and to the calculation of the confining potentials. In Sec. III we characterize the hole states in terms of spatial symmetry and localization in the device, band mixing, and interlevel spacing. The dependence of the Rabi frequencies on the magnetic field (intensity and orientation) and on the geometry of the silicon channel are discussed in Sec. IV and Sec. V, respectively. The conclusions are drawn in Sec. VI, while further details on the method are provided in the Appendices A-D. 

\section{The silicon pMOSFET}

The devices we consider [Fig. \ref{fig:device}(a)] represent scaled pMOSFET versions of those already fabricated in the GlobalFoundries 22nm Fully-depleted Silicon-on-insulator (FDSOI) process \cite{Carter2016a}. The possibility of monolithic integration between qubits and high-fidelity readout circuitry at 2 K in production 22nm FDSOI CMOS technology has been recently demonstrated \cite{Bonen2019a}. The further scaling of the transistor minimum feature size represents a precondition for operating the qubits at high temperatures and control frequencies. 

\begin{figure}[b]
\includegraphics[width=1.\columnwidth]{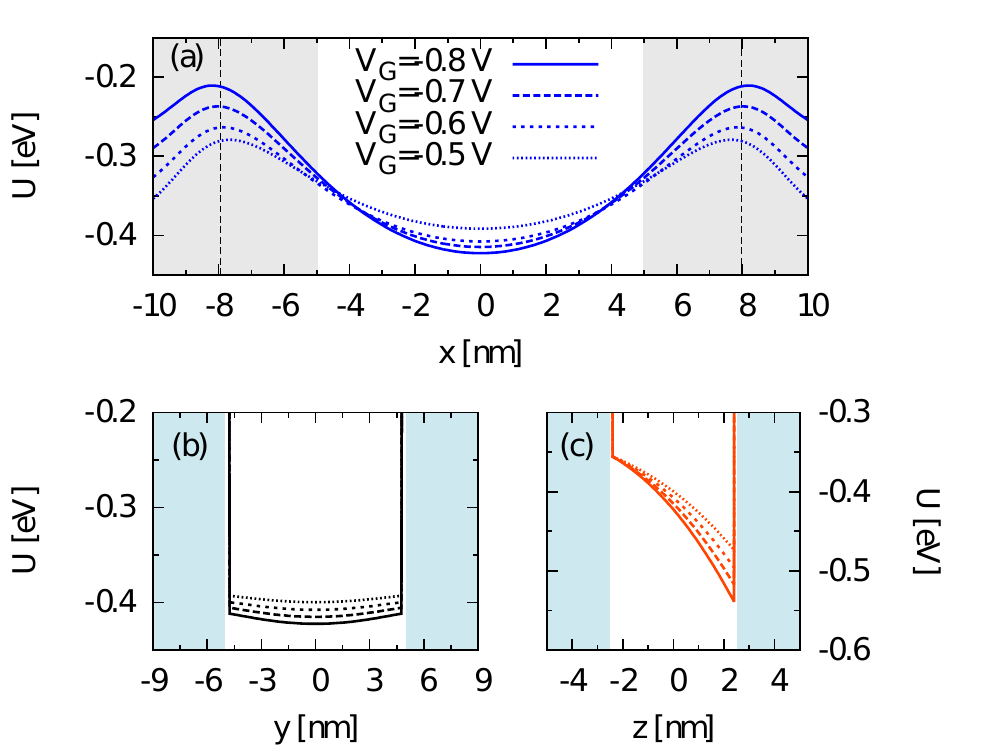}
\caption{Confining potential $U(\boldsymbol{r})$ induced by the top and back gates along the symmetry axes of the silicon channel:  (a) $y\!=\!z\!=\!0\,$nm,  (b) $x\!=\!z\!=\!0\,$nm, and (c) $x\!=\!y\!=\!0\,$nm. In (a) the shaded areas ($|x| \ge 4.98\,$nm) correspond to the source and drain regions, while the dotted lines delimit the region below the nitride spacers ($7.96\,$nm$\,\ge |x| \ge 4.98\,$nm). In (b) and (c) the shaded areas identify the $\text{SiO}_2$ regions ($|y| \ge 5\,$nm and $|z| \ge 2.5\,$nm, respectively). The device temperature in the simulation is $T=2\,$K.}
\label{fig:3DpotVG800}
\end{figure}
The quantum dot is formed in the thin undoped semiconductor film below the top gate. The shallow trench isolation oxides on the two sides of the Si channel ($y$ direction) and the top-gate and buried oxides ($z$ direction) form almost infinite potential barrier wells with dimensions $W=10\,$nm and $H= 5\,$nm, respectively [panel (d)]. The confinement along the $x$ direction results from the combined effect of the top- and back-gate voltages ($V_G$ and $V_B$), of the nitride spacers and of the non-uniform p-type doping in the source/drain (S/D) regions. 

More in detail, the high-$\kappa$ metal gate stacks (length $L_G=$ 10\,nm) consists of 1\,nm SiO\textsubscript{2} interfacial layer and 1.5\,nm HfO\textsubscript{2}; the metal contact on top is simulated by a TiN interface with a 4.57\,eV work function. The stack is separated from the S/D contacts (length $L_s=$ 10\,nm) by 3\,nm Si\textsubscript{3}N\textsubscript{4} spacers [panels (b,c)]. The S/D regions are simulated by realistic non-uniform p-type doping distributions, consisting of a Gaussian profile along the $z$ direction, a uniform profile (with a Gaussian edge) along the $x$ direction, and a peak of the acceptor doping concentration of -10\textsuperscript{20} cm\textsuperscript{-3}. The Si channel (with n-type donor concentration of 10\textsuperscript{15} cm\textsuperscript{-3}) is thus formed between the top gate and the (20\,nm thick) SiO\textsubscript{2} buried oxide, and delimited by the S/D doped regions [see panel (e)]. 

The simulation of this device is carried out by a multiscale approach \cite{ginestra,Vandelli2011a,Padovani2013a,Padovani2017a}. Geometry and composition of the device are estimated from Ref. \onlinecite{Carter2016a}, on the basis of scaling laws. The device is simulated with the simulation software Ginestra\textsuperscript \textregistered. This allows us to determine the three-dimensional potential energy profile that confines the holes and defines the quantum dot, starting from the device geometry and material properties, through a self-consistent description of charge transport (see Appendix A). Crucially, the software Ginestra\textsuperscript \textregistered has the capability to simulate the device physics at the temperature $T=$ 2\,K.

A representative example of confining potentials $U(\boldsymbol{r})$ generated in the pMOS for different values of the top gate potential $V_G$ and a channel width $W=10\,$nm is reported in Fig. \ref{fig:3DpotVG800}. At zero drain-source bias and for a back-gate voltage $V_B= 0.5\,$V, a hole quantum dot is clearly formed in the Si channel, as a combined effect of the band offsets at the Si/SiO$_2$ interfaces and of the electrostatic potential induced by the gates. In particular, the band offset at the Si/Si$\text{O}_2$ interface produces sharp confining barriers in the $y$ and $z$ directions [panels (b,c)]. The applied voltage induces a potential gradient along $z$, which tends to localize the holes close to the top gate. The potential profile is qualitatively different (much smoother) along the channel direction ($x$). Here, the top gate induces a central minimum, separated from the doped source/drain regions by two tunnelling barriers. The depth of the minimum increases for decreasing values of the gate voltage $V_G$, which can be varied within a significant range of values, thus allowing the tuning of the qubit properties (see below). The formation of a well-defined quantum dot in the center of Si channel is obtained for $V_G \lesssim -0.3\,$V. However, in order to remain safely within the required confinement regime, we focus hereafter on more negative values of the gate voltage. Further details on the TCAD simulations are reported in Appendix A.

\begin{figure}[b]
\includegraphics[width=1.\columnwidth]{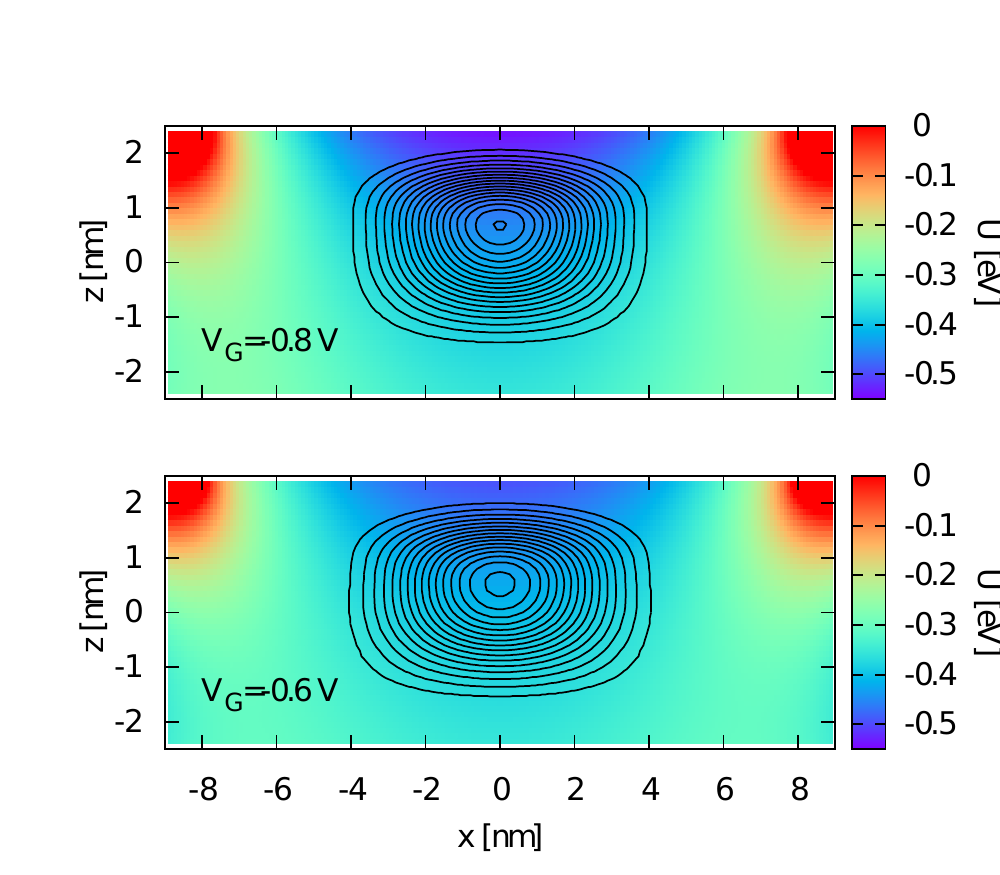}
\caption{Map of the confining potential $U(\boldsymbol{r})$ at $y=0\,$nm, for different values of the gate potential ($V_G=-0.8\,$V and $V_G=-0.6\,$V in the top and bottom panels, respectively). The black contour lines show the charge density $\rho_{1,\xi} (\boldsymbol{r})= |\langle \boldsymbol{r}|1,\xi\rangle|^2$, corresponding to the hole ground states at zero magnetic field, from a value of 0.1 (outer line) to 0.95 (inner line), with an incremental step of 0.05 (all in a.u.).  }
\label{fig:gswf}
\end{figure}
\section{Hole states}

The confining potential $U(\boldsymbol{r})$ generated by the TCAD simulations represents the starting point for the calculation of the hole eigenstates, which is performed by diagonalizing a 6 band L\"uttinger-Kohn envelope function Hamiltonian $H_{\boldsymbol{k} \cdot \boldsymbol{p}}$ \cite{kpbook} (see Appendix B). The hole eigenstates, which at zero magnetic field form degenerate Kramers doublets, are denoted hereafter with:
\begin{eqnarray}
    |m,\xi\rangle 
    &=& \sum_b \sum_{\boldsymbol{n}} c_{m,\xi}^{\boldsymbol{n},b} |\boldsymbol{n},b\rangle .
\end{eqnarray}
Here, the index $m\ge 1$ specifies the doublet and $\xi=\Uparrow,\Downarrow$ the individual eigenstate. The eigenstates are expanded on a basis of band-orbital states specified by the indexes $\mathbf{n}$ and $b$. The band index, $b \equiv (J,M)$, distinguishes between heavy-hole $(3/2,\pm3/2)$, light-hole $(3/2, \pm 1/2)$, and split-off $(1/2, \pm 1/2)$ bands. The orbital index $\mathcal{n} \equiv (n_x, n_y, n_z)$ specifies a set of 3D harmonic oscillator states, on which the envelope functions are expanded.

The qubit initialization and manipulation requires the application of a static magnetic field ${\bf B}=B(\cos\theta\cos\phi,\cos\theta\sin\phi,\sin\theta)$, whose coupling with the confined holes is described by a Hamiltonian $H_{B}$, which includes the Zeeman ($H_{B,Z}$), the paramagnetic ($H_{B,P}$), and the diamagnetic term ($H_{B,D}$)  (see Appendix B). The field removes the degeneracy within each Kramers doublet and thus defines unambiguously the pair of eigenstates $|m,\xi\rangle$, which depend on the field orientation ($\theta$, $\phi$) and intensity ($B$). 

In the remainder of this section, we discuss the properties of the hole states for a prototypical scaled pMOS device with channel width $W=10\,$nm. The depencence of the hole properties on the channel width are discussed in Sec. \ref{sec:dep_on_W}.

\subsection{Quantum dot formation}

The actual formation of a hole quantum dot implies the three-dimensional localization within the Si channel of (at least) the hole ground states $|1,\xi\rangle$ ($\xi = \Uparrow, \Downarrow$), which is indeed achieved in the considered case (Fig. \ref{fig:gswf}).  

In order to quantify the degree of localization for the lowest hole eigenstates, we compute their penetration probabilities in the oxide barriers ($\mathcal{R}_{Ox}$), in the entire source/drain region ($\mathcal{R}_{SD}$), and specifically below the nitride spacers ($\mathcal{R}_{NS} \subset \mathcal{R}_{SD}$). For the generic region $\mathcal{R}$ and eigenstate $|m,\xi\rangle$, the penetration probability reads:
\begin{equation}
p_m(\mathcal{R})=\int_\mathcal{R} d\boldsymbol{r}\, |\langle \boldsymbol{r}|m,\xi\rangle|^2\,.
\end{equation}
These probabilities are independent of $\xi = \Uparrow, \Downarrow$ at zero magnetic field, and display a negligible dependence on ${\bf B}$ in the considered range of field values. 
Their values are at most of the order of $1\%$ for the states belonging to the ground doublet (Table \ref{tab:unique}), and only slightly larger for the first eight excited states (not reported here). Most of the leakage takes place along the $x$ direction, towards the source and drain, in particular in the region below the nitride spacers. 
The hole confinement is stronger in the $y$ and $z$ directions, where the large band offset between Si and SiO$_2$ limits the leakage probabilities to less than $0.2\%$. The comparison between the probabilities corresponding to different values of $V_G$ shows a clear dependence of $p(\mathcal{R}_{SD})$ and $p(\mathcal{R}_{NS})$ on the top-gate voltage, which has instead a limited impact on the $p(\mathcal{R}_{Ox})$.

The energy separation $\Delta$ between the ground and the first excited doublet quantifies the strength of the quantum confinement in the dot and thus provides an estimate of the highest temperature at which its effects can be observed. Here, such gap is given by $\Delta = 6.58\, (6.95)\,$meV at $V_G=-0.8\, (-0.6)\,$V, which corresponds to a temperature of $T = \Delta / k_B = 76.358\, (80.651)\,$K. Overall, this shows the possibility of obtaining well defined quantum dots within the present geometry, with strongly confined hole states. 


\subsection{Band mixing}

The mixing between the subbands affects a number of relevant properties of the hole qubit, and represents a prerequisite for its manipulation by means of electric fields \cite{Venitucci2018a}. In order to quantify such mixing, we refer in the following to the probabilities
\begin{align}
P_{m,\xi}^{\chi} = \sum_{b \in \chi} p_{m,\xi}^{b} = \sum_{b \in \chi} \left( \sum_{\boldsymbol{n}} | c_{m,\xi}^{\boldsymbol{n},b} |^2\right)\,,
\end{align}
which give the weights, for the hole eigenstate $|m,\xi\rangle$, of the heavy-hole ($\chi\! =\! hh$), light-hole ($\chi\! =\! lh$),
and split-off bands ($\chi\! =\! so$). In the weak-field limit, where the magnetic field does not induce a significant mixing between
states belonging to different doublets, one has that $ P_{m,\Uparrow}^{\chi} = P_{m,\Downarrow}^{\chi} $, and these probabilities are independent of the field orientation (while this is not the case for the individual-band probabilities $p_{m,\xi}^{b}$). 

In all the considered cases, the ground state presents a largely heavy-hole character, but also a significant occupation of the light-hole subbands, corresponding to values of $p_{1,\xi}^{lh}$ larger than $10\%$, as reported in Table \ref{tab:unique}. This results from the fact that the strength of the vertical ($z$) confinement is comparable to that in the $x$ and (especially) $y$ directions (see Fig. \ref{fig:3DpotVG800}).  
\begin{table}[t]
\renewcommand{\arraystretch}{1.2}
\begin{tabular}{| c | c c c c|}
\hline
 $V_G$ [V] & $p(\mathcal{R}_{SD}) [$\%$]$ & $p(\mathcal{R}_{NS})$ [$\%$] &  $p(\mathcal{R}^y_{Ox})$ [$\%$] &  $p(\mathcal{R}^z_{Ox})$ [$\%$]\\
\hline
-0.8 & 0.91 & 0.77 & 0.01 & 0.19 \\
-0.6 & 1.42 & 1.23 & 0.01 & 0.18 \\
\hline
\hline
 $V_G$ [V] & $p_{1,\xi}^{hh}$ [$\%$] & $p_{1,\xi}^{lh}$ [$\%$] & $p_{1,\xi}^{so}$ [$\%$] &  \\
\hline
-0.8 & 86.44 & 11.57 & 1.99 &  \\
-0.6 & 83.14 & 14.56 & 2.30 &  \\
\hline
\hline
 $V_G$ [V] & $\langle\sigma_{yz}\rangle$ & $\langle\sigma_{zx}\rangle$ & $\langle\sigma_{xy}\rangle$ & $\langle\sigma_{r}\rangle$ \\
\hline
-0.8 & 0.9655 & 0.9670 & 0.8548 & 0.8870 \\
-0.6 & 0.9550 & 0.9692 & 0.8894 & 0.9289 \\
\hline
 \end{tabular}
 \caption{Characterization of the hole ground state in terms of wave-function localization, symmetries, and band occupation (upper, middle, and lower panels, respectively).
 In particular, the penetration probabilities refer to the regions: of the source-drain ($\mathcal{R}_{SD}$, corresponding to $10\,$nm$\,\ge |x| \ge 4.98\,$nm), below the nitride spacers ($\mathcal{R}_{NS}$, $7.96\,$nm$\,\ge |x| \ge 4.98\,$nm), of the oxides that confine the holes along the $y$ ($\mathcal{R}^y_{Ox}$, $8.4\,$nm$\,\ge |y| \ge 5.04\,$nm) and $z$ directions ($\mathcal{R}^z_{ox}$, $3.5\,$nm$\,\ge |z| \ge 2.5\,$nm).} 
 \label{tab:unique}
\end{table}
The split-off band presents a limited weight ($p_{1,\xi}^{so} \lesssim 2\%$). However, it provides an indirect coupling between the heavy- and light-hole subbands, which significantly affects their occupation probabilities. Its inclusion in the calculation is thus required in order to get an accurate description of the hole eigenstates. In particular, this applies to the Rabi frequencies, which are extremely sensitive to small changes in the mixing between the heavy and light holes (see Sec. \ref{sec:lrfr}). 

\subsection{Wave function symmetry}

One additional feature plays a role in the manipulation of the hole-spin qubit, namely the symmetry of the ground-state wave function. As shown in Fig. \ref{fig:gswf}, the charge density in the ground state is asymmetric in the $z$ direction (the hole is attracted towards the top gate) and approximately symmetric along the $x$ and $y$ axes. A more precise and quantitative characterization of the eigenstate symmetry is obtained by computing the expectation values of the operators $\sigma_{\alpha\beta}$, which implement a reflection of the wave function envelope in the $\alpha\beta$ plane ($\alpha\beta=xy,yz,zx$), and $\sigma_r$, which implements a spatial inversion ($\boldsymbol{r}\!\rightarrow\! -\boldsymbol{r}$). Their expectation values can be directly deduced from the symmetry properties of the basis states, and are given by:
\begin{align}
&\langle m,\xi |\sigma_{\alpha\beta}|m,\xi\rangle = 
\sum_{b}\sum_{\boldsymbol{n}} |c_{m,\xi}^{b,\boldsymbol{n}}|^2 (-1)^{n_\gamma}\,
\label{eq:symmetry1}
\end{align}
(where $\gamma \neq \alpha,\beta$) and
\begin{align}
&\langle m,\xi |\sigma_r|m,\xi\rangle \!=\! 
\sum_{b}\sum_{\boldsymbol{n}} |c_{m,\xi}^{b,\boldsymbol{n}}|^2 (-1)^{n_x+n_y+n_z}\,.
\label{eq:symmetry2}
\end{align}
The expectation values of the symmetry operators introduced above are between $-1$ and $+1$, where the extremal values denote that the hole eigenstate has a complete odd or even symmetry, respectively. An intermediate value signals that the hole eigenstate is an admixture of odd and even orbital states. 

The ground state is only approximately symmetric in all directions (Table \ref{tab:unique}). In fact, non-negligible deviations of the expectation values from 1 are obtained, especially for $\sigma_{xy}$ and $\sigma_r$. The value of $\langle\sigma_{xy}\rangle$ decreases for decreasing values of $V_G$, thus reflecting the relative relevance in the $z$-confinement of the oxide barriers (which are symmetric) and of the gate voltage (approximately antisymmetric). The smaller deviations from 1 of $\langle\sigma_{yz}\rangle$ and $\langle\sigma_{zx}\rangle$ result from the light-hole component of the ground state, and are due to the band-mixing (off-diagonal) terms in $H_{\boldsymbol{k} \cdot \boldsymbol{p}}$. 
In general, a lower degree of symmetry of the hole ground state and/or in the spatial profile of the manipulating electric fields represents a potential advantage in terms of qubit manipulation capabilities, for it reduces the symmetry-related selection rules that have to be fulfilled in order to induce the desired transitions \cite{Venitucci2019a}.

\section{Larmor and Rabi frequencies\label{sec:lrfr}}

The static magnetic field $\boldsymbol{B}$ plays a fundamental role in the initialization and manipulation of the hole-spin qubit. 
In fact, it opens a gap between the two (otherwise degenerate) ground states $|1,\Uparrow\rangle$ and $|1,\Downarrow\rangle$, and thus determines the Larmor frequency, 
\begin{align}
&f_L=\frac{1}{h}\left( E_{1,\Uparrow} - E_{1,\Downarrow} \right) \,,
\label{eq:flar}
\end{align}
at which the spin manipulation has to be performed by means of oscillating fields. Besides, it affects the band mixing in the ground states, and creates the possibility of inducing transitions between them through electric fields.
Within the present geometry, the spin-qubit manipulation is performed by means of time-dependent voltages applied to the top gate (Fig. \ref{fig:device}). In order to assess the possibility of implementing arbitrary single-qubit gates, we compute hereafter the Rabi frequencies, corresponding to unitary transformations that are either diagonal or off-diagonal in the eigenstate basis. 

Transverse rotations are here identified with unitary transformations $\exp(-i\varphi_X\sigma_X/2)$, where $\sigma_X$ is the first Pauli matrix (expressed in the $\{|1,\Downarrow\rangle,|1,\Uparrow\rangle\}$ basis), and allow transitions between the qubit eigenstates. Physically, such rotations are implemented by a resonant ac voltage $\delta V_G \cos (f_L t)$, and are characterized by the Rabi frequency:
\begin{align}
&f_R^{X}=\frac{1}{h}\left|\langle 1,\Uparrow\!|\delta U|1,\Downarrow \rangle \right| \,,
\label{eq:freXa}
\end{align}
where $\delta U (\boldsymbol{r}) = U_{V_G+\delta V_G}(\boldsymbol{r}) - U_{V_G}(\boldsymbol{r})$ is the difference between the confining potentials corresponding to the gate voltages $V_G+\delta V_G$ and $V_G$. 

Longitudinal rotations correspond to unitary transformations $\exp(-i\varphi_Z\sigma_Z/2)$, and modify the relative phase of the qubit eigenstates. Physically, this is obtained by modulating the energy gap between the two basis states through a dc voltage $\delta V_G$. The corresponding Rabi frequency is given by:
\begin{align}
&f_R^{Z}=\frac{1}{2h}\left| \langle 1,\Uparrow\! |\delta U|1,\Uparrow\rangle- \langle1,\Downarrow\!| \delta U|1,\Downarrow\rangle \right| \, .
\label{eq:freZa} 
\end{align}

In general, it is not possible to selectively turn on, through the time-dependent voltage, either the diagonal ($\xi = \xi'$) or the off-diagonal elements ($\xi\neq\xi'$) of the differential potential matrix $ \langle 1,\xi |\delta U |1,\xi'\rangle$. However, the modulation of the diagonal elements can be neglected while considering rotations around the $X$ axis, because it takes place at a frequency $f_L \gg f_R^Z $ and thus averages to zero during the ac voltage pulse (in the spirit of the rotating wave approximation). Analogously, the off-diagonal elements of $\delta U$ don't affect the rotations around the $Z$ axis, because they are much smaller than the difference between the diagonal ones ($f_L \gg f_R^X$). 

\begin{figure}[t]
\includegraphics[width=1.\columnwidth]{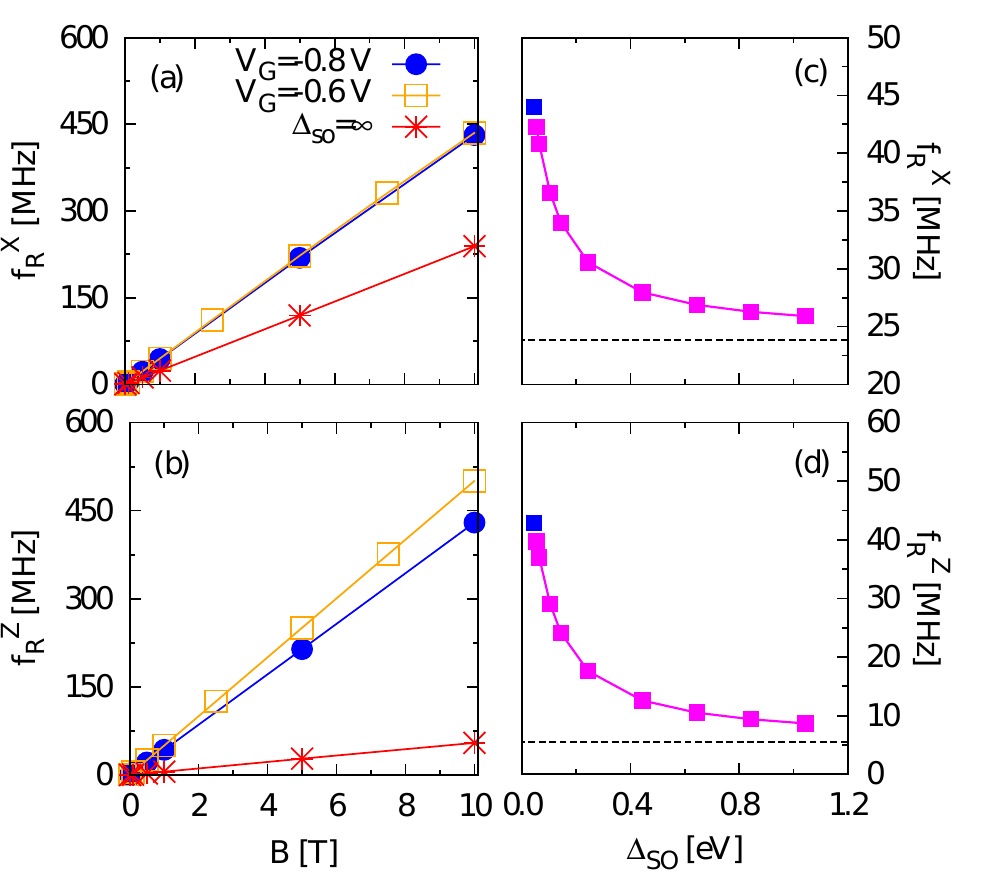}
\caption{Rabi frequencies (a) $f_R^X$ and (b) $f_R^Z$ as a function of the magnetic field intensity $B$ ($\theta=45^\circ$ and $\phi=0^\circ$).  
Rabi frequencies (c) $f_R^X$ and (d) $f_R^Z$ as a function of $\Delta_{SO}$, ranging from $0.044\,$eV (its actual value in Si, blue squares) to $1\,$eV ($B=1\,$T, $V_G=-0.8\,$V). The limit $\Delta_{SO}=\infty$ (dotted lines) corresponds to calculations performed without including the split-off band.} 
\label{fig:rabifW10}
\end{figure}

\subsection{Dependence of the Rabi frequencies on the magnetic-field intensity}

The electric field cannot directly couple two states that are conjugated by time-reversal symmetry, as is the case for $|1,\Uparrow\rangle$ and $|1,\Downarrow\rangle$ in the zero-field limit (see Appendix C). In order for $f^X_R$ to take a finite value, the static 
magnetic field must thus mix states belonging to different Kramers doublets \cite{Venitucci2018a}. Such mixing is induced by the different components of the 
magnetic-field Hamiltonian, and results in a linear dependence on the field intensity for both $f^X_R$ and $f^Z_R$. Here, in particular, both 
Rabi frequencies increase with a rate of $45\,$MHz$/$T at $B=1\,$T, for a non-optimal orientation of the static magnetic field [Fig.~\ref{fig:rabifW10} (a,b)]. 
This demonstrates the possibility of efficiently manipulating the hole-spin qubits within the planar FDSOI geometry and, more specifically, in the Si scaled pMOSFET structure that we are considering. 

In the present geometry, all the introduced magnetic-Hamiltonian terms and bands contribute significantly to the values of the Rabi frequencies. In fact, the values of both $f_R^X$ and $f_R^Z$ obtained with the inclusion of the split-off band are at least twice as large as those obtained within a four-band approach. The four-band case can be considered as the limit of the six-band case for $\Delta_{SO} \rightarrow \infty$; by artificially increasing $\Delta_{SO}$ in the six-band calculations, we see that the Rabi frequencies tend to those obtained from the four-band calculations [Fig. 4 (c,d)].
The difference between the Rabi frequencies does not result from the direct contribution of the split-off band to the Rabi frequencies, but rather from the enhanced mixing that this band induces between light and heavy holes (Appendix C).
The contribution that comes from the sum of the paramagnetic and diamagnetic terms in the Hamiltonian is comparable, for both $f_R^X$ and $f_R^Z$, to the value obtained with the Zeeman term alone (not shown). 

\begin{figure}
\includegraphics[width=1.\columnwidth]{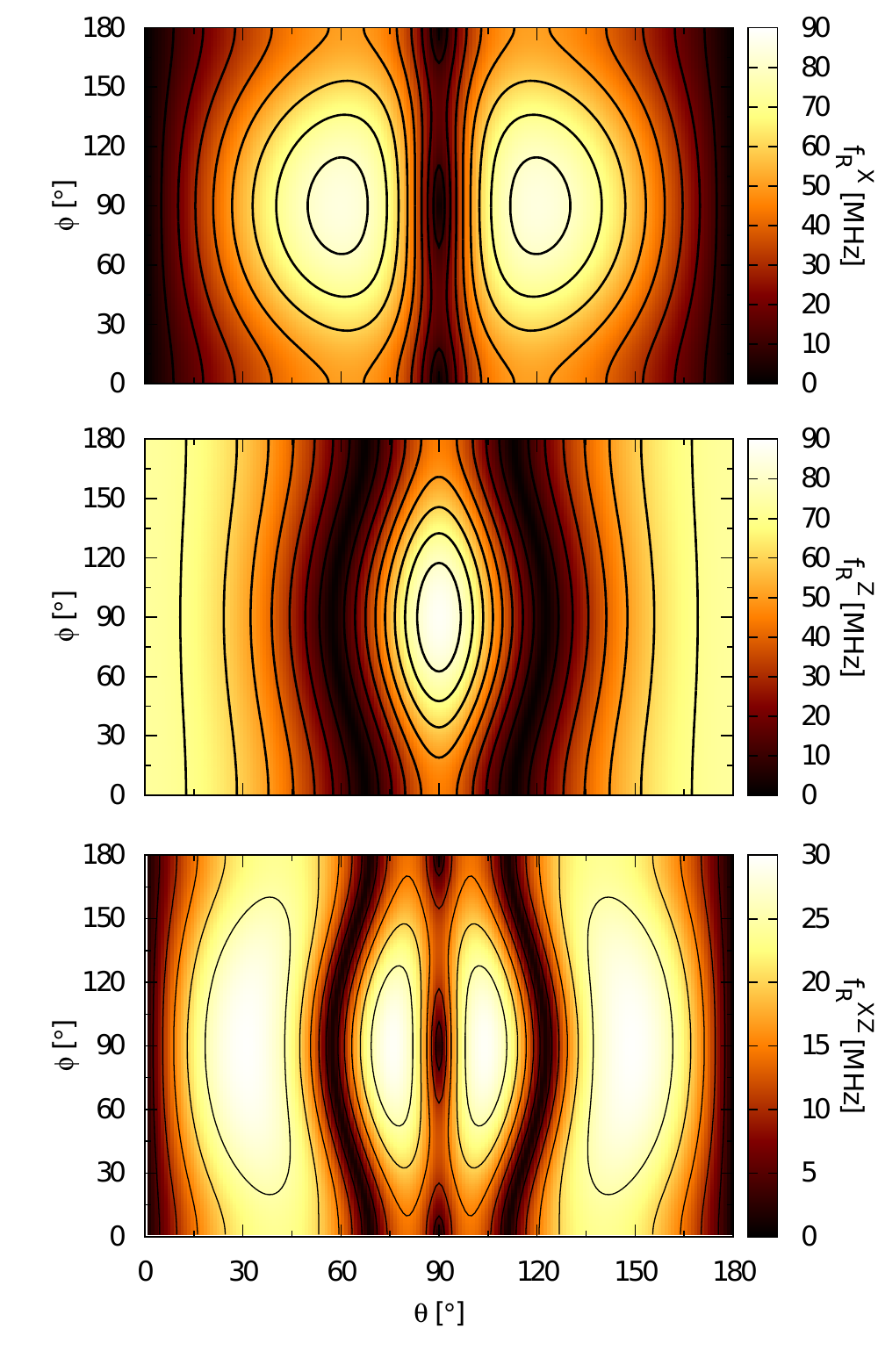}
\caption{Dependence of the Rabi frequencies $f_R^{X}$ (upper panel) and $f_R^{X}$ (middle), and of $f_R^{XZ}$ (lower panel) on the orientation of the static magnetic field. The plotted quantities are computed by means of the $g$-matrix formalism ($V_G=-0.8\,$V, $\delta V_G=10\,$mV, and $B=1\,$T). }
\label{fig:rabimap10}
\end{figure}

\subsection{Dependence of the Larmor and Rabi frequencies on the magnetic-field orientation}

The multiband calculations show that the Larmor frequency varies linearly with the field intensity. In such a regime, the single hole can be formally described as a two-level system within the so-called {\it g-matrix formalism} \cite{Kato2003a,Crippa2018a}, which we use in the following to illustrate the dependence of the qubit properties on the field orientation, and to interpolate the results of the $\bf{k}\cdot\bf{p}$ calculations. In the present case, the $g$ matrix can be approximately written in a diagonal form (see Appendix D), and the effective qubit Hamiltonian reads:
\begin{align}
\mathcal{H}=\frac{1}{2}\mu_B(g_{x} B_{x} \sigma_X + g_{y} B_{y} \sigma_Y + g_{z} B_{z}\sigma_Z)\,.
\label{eq:Hgdiag}
\end{align}
Here, the coordinate axes $x$, $y$, and $z$ can be identified respectively with the directions $[110]$, $[\overline{1}10]$, and $[001]$, $|\xi\rangle\equiv |1,\xi\rangle$ ($\xi=\Uparrow,\Downarrow$) are with the hole eigenstates  obtained for a magnetic field along the $z$ direction, and the $g$ factors depend on the gate voltage $V_G$. From the above qubit Hamiltonian, it follows that the Larmor frequency can be expressed as a function of the $g$ factors, according to the expression:
\begin{align}
f_L=&\frac{\mu_B B}{h}\sqrt{(g_x b_x)^2 \!+\! (g_y b_y)^2 \!+\! (g_z b_z)^2} \equiv \frac{\mu_B B}{h} g^* \,,
\label{eq:larmoran}
\end{align}
where $g^*$ is the effective $g$ factor and 
$\boldsymbol{b}\!=\!\textbf{B}/B=(b_x,b_y,b_z) = (\sin\theta\cos\phi,\sin\theta\sin\phi,\cos\theta)$ is the unit vector  that defines the field orientation. The Larmor frequency is thus maximum (minimum) for a magnetic field aligned in the direction of the largest (smallest) $g_\alpha$. 

The degree of anisotropy of the $g$ tensor results from the interplay of band mixing and magnetic Hamiltonian, including all the magnetic terms. In fact, while purely $hh$ states in the presence of a  Zeeman term alone would give $g_x\!=\!g_y\!=\!0$ and $g_z\!=\!-6\kappa\! =\! 2.57$, here we find that $g_z$ is almost twice that value, and $g_x$ and $g_y$ differ from zero (Table \ref{tab:Geigen}). The predominant $hh$ character of the hole ground state and the presence of a small $lh$ contribution (see Table \ref{tab:unique}) characterize all the considered values of the top-gate voltage, which however allows a certain degree of tunability of the band mixing, and thus of the $g_\alpha$. In fact, decreasing values of $V_G$ increase the strength of the vertical ($z$) confinement (relative to that in the $xy$ plane), and thus the weight of the $hh$ subbands. This, in turn, results in an increase of $g_z$ and in a decrease of $g_x$ and $g_y$.  Besides, we note that the Larmor frequency is mostly determined by the Zeeman Hamiltonian for a predominant in-plane component of the magnetic field, while the paramagnetic Hamiltonian contributes mostly when ${\bf B}$ aligns to the $z$ direction.

The Rabi frequencies can be derived from the $g$-tensor and its derivative with respect to the gate potential, $\hat{g}'=\partial \hat{g} / \partial V_G $ \cite{Venitucci2018a,Ares2013a,Voisin2015a}. In particular, the expressions of $f_R^{X}$ and $f_R^{X}$  read:
\begin{align}
&f_R^{X}=
\frac{\mu_B B \delta V_{G}}{2hg^*}|(\hat{g}\boldsymbol{b}) \times (\hat{g}'\boldsymbol{b})| 
\label{eq:freX}
\end{align}
and  
\begin{align}
&f_R^{Z}=
\frac{\mu_B B \delta V_{G}}{2 h g^*}|(\hat{g}\boldsymbol{b})\cdot (\hat{g}'\boldsymbol{b})|\,.
\label{eq:freZ} 
\end{align}
Our calculations show that the perturbation of the confining potential associated with the time-dependent voltages essentially preserves the symmetry of the unperturbed Hamiltonian. As a result, $\hat{g}'$ shares with $\hat{g}$ the principal axes, and the derivative of the $g$ matrix with respect to $V_G$ can be reduced to that of the principal $g$ factors, whose values are reported in Table \ref{tab:Geigen}. Formally, this simplifies the expressions of the Rabi frequencies, since $\hat{g}'\boldsymbol{b}=\sum_{\alpha=x,y,z} g_\alpha' b_\alpha$. From a physical point of view, it implies that the hole manipulation can only occur through the $g$-tensor magnetic resonance, while the iso-Zeeman contribution to the spin-electric coupling vanishes.

\begin{table}[t]
\renewcommand{\arraystretch}{1.2}
\begin{tabular}{| c |c c c | c c c |}
\hline
$V_G$ [V] & $g_x$ & $g_y$ & $g_z$ & $g_x'$ & $g_y'$ & $g_z'$ \\
\hline
-0.6 & 1.342 & 1.700 & 4.074 & 0.553 & 1.192 & -1.209 \\
-0.8 & 1.222 & 1.451 & 4.299 & 0.639 & 1.261 & -1.037 \\
\hline
\hline
$V_G\,$[V] & $\theta_{R,max}^X$ & $f_{R,max}^X$ & & $\theta_{R,min}^Z$ & $f_{R,max}^Z$ & \\  
\hline
-0.6 & $57.1^\circ$ & $83.8\,$MHz & & $57.3^\circ$ & $83.4\,$MHz & \\  
-0.8 & $59.9^\circ$ & $84.3\,$MHz & & $57.3^\circ$ & $88.3\,$MHz & \\  
\hline
\end{tabular}
\caption{Upper panel: Values of the principal $g$ factors computed for two different values of the top-gate voltage $V_G$. Lower panel: Maxima of the Rabi frequency $f_R^X$ ($f_R^Z$), and values of the angle $\theta$ that correspond to their maxima (minima). The magnetic-field intensity is $B=1\,$T.}
\label{tab:Geigen}
\end{table}

The dependence of the Rabi frequencies on the orientation of the magnetic field can be derived from the above equations, once the terms $g_\alpha$ and $g_\alpha'$ have been obtained through the $\bf{k}\cdot\bf{p}$ calculations. We start by noting that the Rabi frequency $f_R^X$ necessarily vanishes if the magnetic field is aligned along one of the principal axes. In fact, in this case, the vector $\hat{g}\boldsymbol{b}$ is parallel to $\hat{g}'\boldsymbol{b}$, and the cross product in Eq.~(\ref{eq:freX}) vanishes. 
The overall dependence of $f_R^X$ on $\theta$ and $\phi$ is reported in Fig.~\ref{fig:rabimap10}, and is qualitatively similar to that
obtained in different geometries \cite{Venitucci2018a}. The maxima of the Rabi frequency are contained in the $yz$ plane ($\phi = 90^\circ$), for which one can derive an analytical expression (Appendix D). The optimal field orientation is given by \cite{Ares2013a}:
\begin{align}
\tan\theta_{R,max}^{X}=\pm\left(\sqrt{\left|g_z/g_y\right|}\right)\,.
\label{eq:thetamax}
\end{align}
The corresponding maximum of $f_R^X$ is given by
\begin{align}
f_{R}^{X,max}=\frac{\mu_B B V_{ac}}{2h}\frac{\left|g_z g_y'- g_y g_z'\right|}{|g_z|+|g_y|}\,.
\label{eq:frmax}
\end{align}
The values of $\theta_{R,max}^{X}$ and $f^X_{R,max}$ obtained in the present geometry are reported in Table \ref{tab:Geigen}.
The field orientation along the $xz$ planes is in general less favorable. In any case, the expressions of the optimal angle and of the maximum for $\phi=0^\circ$ are obtained respectively from Eq.~(\ref{eq:thetamax}) and Eq.~(\ref{eq:frmax}) by replacing $y$ with $x$ in the subscripts. 

\begin{figure}[b]
\includegraphics[width=1.\columnwidth]{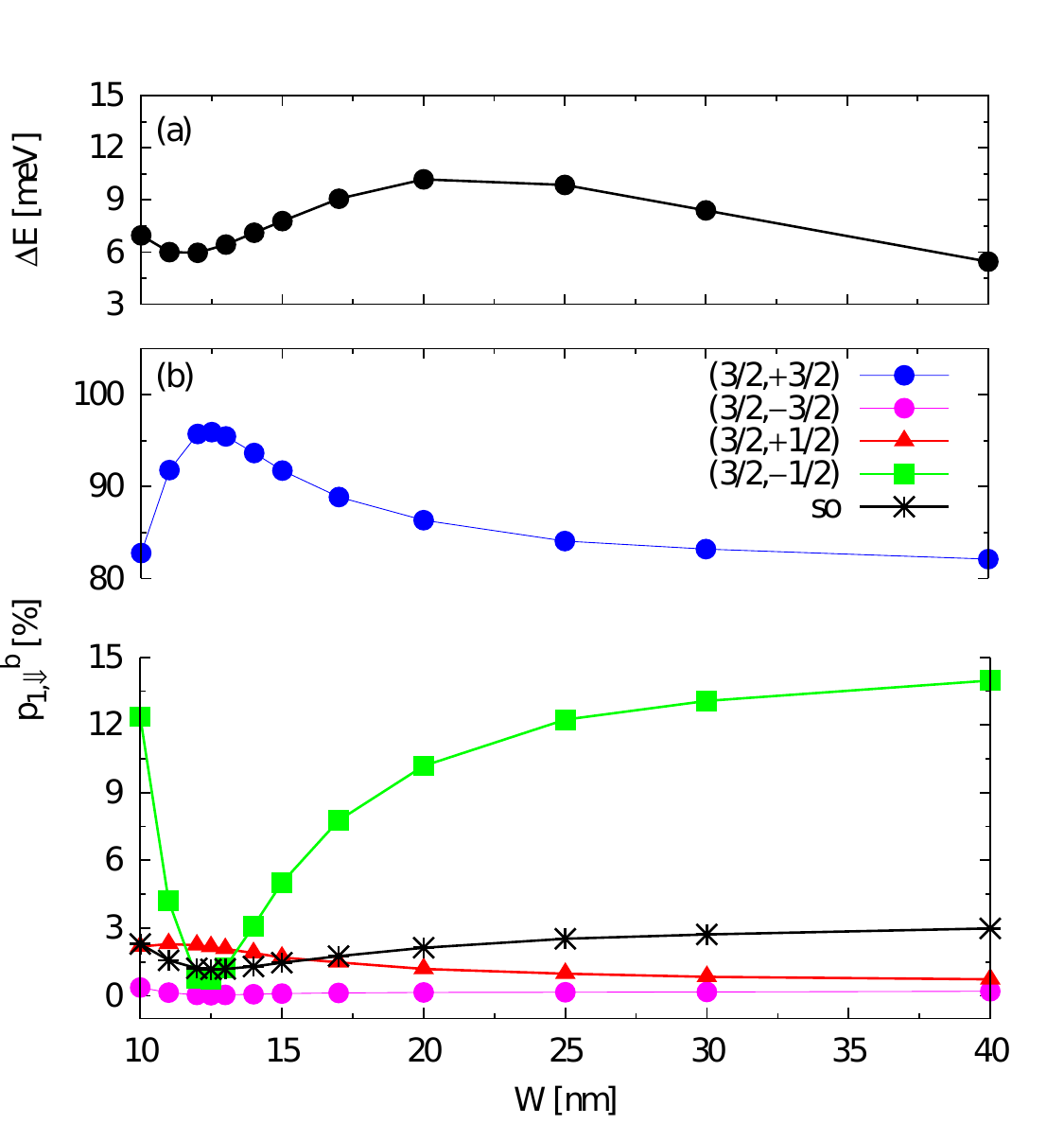}
\caption{(a) Energy splitting between the ground and first excited doublets as a function of the channel width $W$, at zero magnetic field and for $V_G=-0.6\,$V. Occupation probabilities of the (b) $lh$ band, (c) $hh$ and $so$ bands in the ground state in the weak-field limit ($B=10^{-2}\,$T, with $\theta=0^\circ$).}
\label{fig:deltaEW}
\end{figure}
The dependence of the Rabi frequency $f_R^Z$ on the magnetic field orientation is significantly different from that of $f_R^X$. In fact, the field orientations that maximize $f^X_R$ yield vanishing values of $f_Z^R$, and viceversa (Fig.~\ref{fig:rabimap10}). More specifically, as can be readily seen from Eq.~\ref{eq:freZ}, the optimal field direction coincides with the coordinate axis along which the derivative of the $g$ factor is the largest:: 
\begin{align}
f_{R,max}^{Z}=&\frac{\mu_BBV_{ac}}{2h} \max_{\alpha\in\{x,y,z\}} |g_\alpha'|\,.
\label{eq:frmaz}
\end{align}
In the present device geometry, this can either coincide with $y$ or $z$, depending on the top-gate voltage (Table \ref{tab:Geigen}). In any case, the optimal value of $\phi$ is $90^\circ$, as in the case of $f_R^X$.
Given the analytical expression for the dependence of $f_{R}^{Z}$ on the angle $\theta$ (Appendix D), one can also derive a simple expression for the angle at which $f^Z_R$ vanishes, namely
\begin{align}
\tan\theta_{R,min}^{Z}=\pm\sqrt{\left|g_zg_z'/g_yg_y'\right|}\,.
\label{eq:thetamaz}
\end{align}
As in the case of $f_R^X$, the expression for $\phi=0^\circ$ ($xz$ plane) can be derived from the one above by replacing $y$ with $x$.

Overall, the optimal orientation of the magnetic field is one that allows the implementation of both $X$ and $Z$ rotations at a reasonably high Rabi frequency. Given that $f_R^X$ and $f_R^Z$ clearly cannot be maximized by the same field orientation, one needs to identify a trade-off by maximizing a suitable figure of merit, which accounts for both rotations. A simple and yet significant option is represented by the frequency
\begin{equation}
    f_R^{XZ} = \left( \frac{1}{f_R^X} + \frac{1}{f_R^Z} \right)^{-1} = \frac{f_R^X f_R^Z}{f_R^X + f_R^Z} .
\end{equation}
If one identifies $1/f_R^X$ and $1/f_R^Z$ with the duration of respectively $X$ and $Z$ rotations, then $1/f_R^{XZ}$ corresponds to the time required for sequentially implementing the two operations.
The full dependence of $f_R^{XZ}$ on the magnetic field orientation is reported in the bottom panel of Fig.~\ref{fig:rabimap10}. In general, its maximization is obtained for a magnetic field that lies in the $yz$ plane, for values of $\theta$ between the ones that maximize $f_{R}^{X}$ and $f_{R}^{Z}$. 
In spite of the different angular dependencies of the two Rabi frequencies, an overall optimization can thus lead to a satisfactory trade-off. As detailed in the following section, the optimal orientation varies as a function of the channel width. 

\section{Effect of the channel width\label{sec:dep_on_W}}

The size of the devices, and specifically that of the Si channel, plays a crucial role in many respects. Smaller devices have the potential to induce a stronger quantum confinement. This results in larger energy gaps between the Kramers doublets at zero field, and thus allows in principle to use larger magnetic fields, to work at higher Larmor frequencies, and possibly at higher temperatures. However, a change in the quantum confinement also affects the band mixing within the hole eigenstates, and this unavoidably affects all the properties that have been discussed so far. In this respect, a relevant role is also played by the relative strength of the spatial confinement along the different directions. Hereafter, we focus on this latter aspect, and specifically consider the dependence of the hole properties on the width $W$ of the Si channel, a parameter that can be varied in the present pMOSFETs. 

\subsection{Energy gap and band mixing}

We start by considering the dependence of the band mixing and of the energy gaps on the channel width $W$, which we vary from $10$ to $40\,$nm. The gap $\Delta E$ between the ground  ($|1,\xi\rangle$) and first excited ($|2,\xi\rangle$) doublets varies non-monotonically with $W$, but remains above the value of $5\,$meV, corresponding to a temperature of about $60\,$K, in the whole considered range [Fig. \ref{fig:deltaEW}(a)]. The two turning points in such dependence correspond to transitions or avoided level crossings involving the ground or the first excited doublets. 
\begin{figure}[t]
\includegraphics[width=1.\columnwidth]{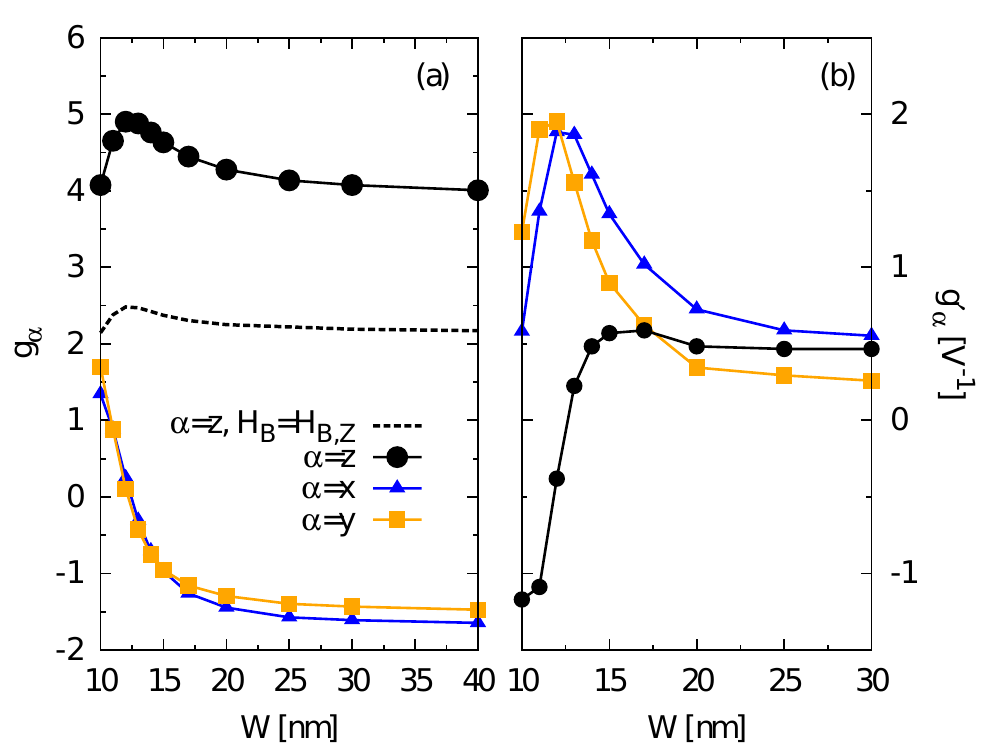}
\caption{Dependence on the channel width $W$ of (a) the $g$ factors and (b) their derivatives with respect to $V_G$, for $V_G=-0.6\,$V. The dotted line in panel (a) gives the values of $g_z$ obtained by neglecting the paramagnetic and diamagnetic contributions in the Hamiltonian ($H_B = H_{B,Z}$). }
\label{fig:ggpW}
\end{figure}
In fact, the minimum (at $W_a\approx 12\,$nm) is coincides with a non-monotonic variation of the band-occupation probabilities for the ground state [Fig. \ref{fig:deltaEW}(b,c)]: the occupation probability $p_{hh}^{1,\xi}$ reaches its maximum value, while it monotonically decreases for higher values of $W$. The opposite occurs with $p_{lh}^{1,\xi}$. The maximum of $\Delta E$ (for $W_b\approx 20\,$nm) coincides instead with a transition in the excited doublet, which is reflected in a change in the expectation value of $\sigma_{yz}$ (not shown).

\begin{figure}[t]
    \centering
    \includegraphics[width=1.\columnwidth]{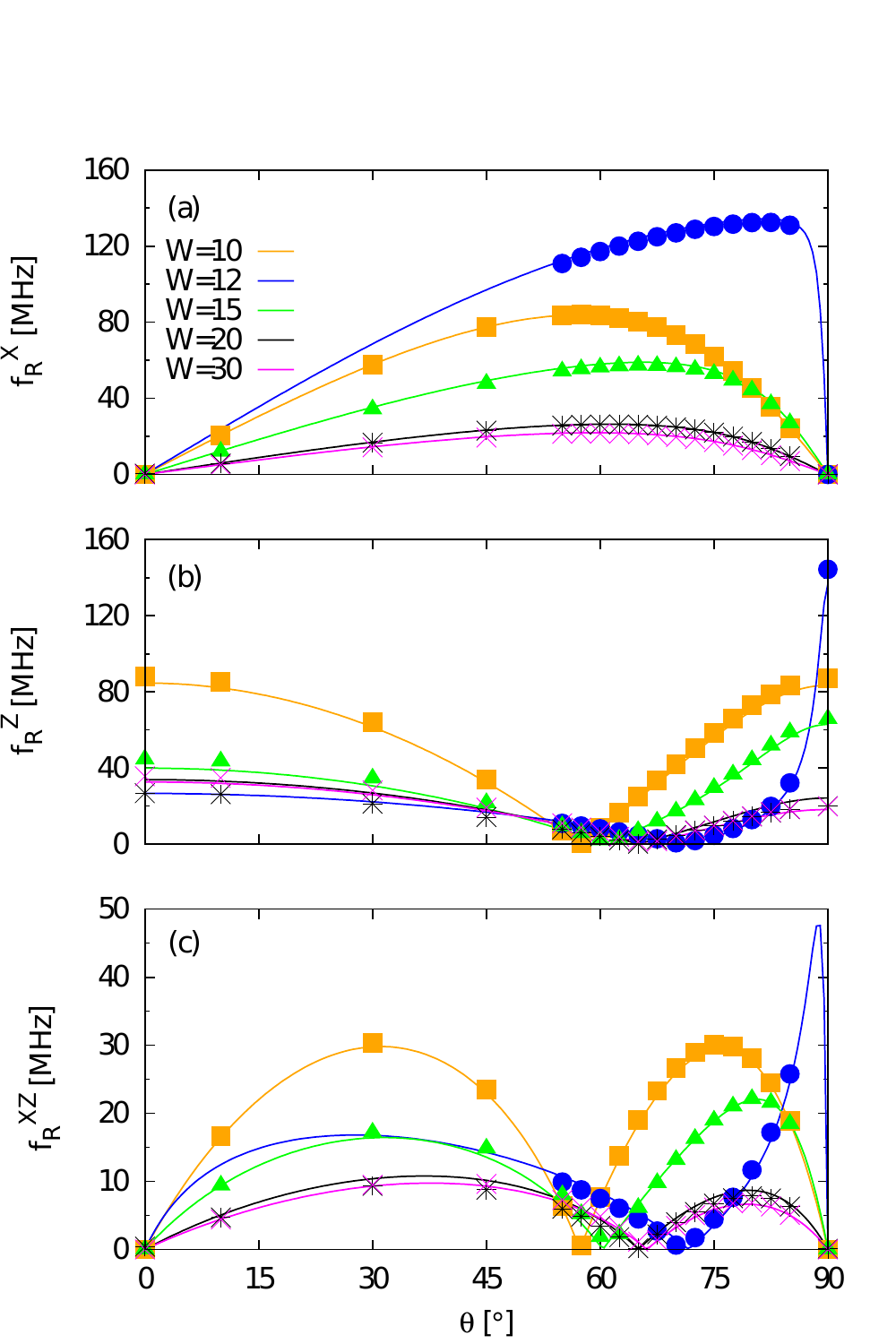}
    \caption{Dependence on the magnetic field orientation of the Rabi frequencies (a) $f_R^X$, (b) $f_R^Z$, and (c) $f_R^{XZ}$ for different values of the channel width $W$, for $V_G=-0.6\,$V, $\delta V_G=10\,$mV, $B=1\,$T, and $\phi=90^{\circ}$. The plots include both the results derived from the $g$-matrix approach (solid lines) and the numerical results obtained directly from the $\bf{k}\cdot\bf{p}$ calculations (symbols).}
    \label{fig:fomW}
\end{figure}
\subsection{Larmor and Rabi frequencies}

The dependence of the band mixing on the channel width is reflected on the $g$ factors, and thus on the Larmor and Rabi frequencies, which can be expressed in terms of the quantities $g_\alpha$ and $g_\alpha'$. Strong variations of all these parameters show up at $W = W_a$ (Fig.~\ref{fig:ggpW}). In particular, the relative change between $g_z$ and $g_y$ result in strong variations of the optical angle $\theta^X_{R,max}$ (see Eq. \ref{eq:thetamax}), which approaches the value of $90^\circ$ where $g_y$ undergoes a sign change. Also, the optimal field orientation for the $Z$ rotations varies from one in-plane axis to the other, due to the different dependence of $g_x'$ and $g_y'$ on the channel width [panel (b)].

The channel width significantly affects also the dependence of the Rabi frequencies on the magnetic field orientation (Fig.~\ref{fig:fomW}). The values of all the maxima ($f_{R,max}^X$, $f_{R,max}^Z$, $f_{R,max}^{XZ}$) display a non-monotonic dependence on $W$, and present a maximum for $W=W_a$, i.e. for a channel width that is slightly larger than its length $L_Q=10\,$nm. As the width approaches the value of $W_a$, the optimal angle $\theta_{R,max}^X$ approaches $90^\circ$, and so do $\theta_{R,max}^Z$ and $\theta_{R,max}^{XZ}$. Therefore, unlike what happens at larger and smaller values of $W$, the field orientations that allow large values of one Rabi frequency don't correspond to vanishingly small values of the other one.  

The present analysis leaves aside the role of decoherence, whose specific features in these particular devices are still unknown. In particular, the low-frequency electrical noise might cause dephasing by inducing random modulations of the energy gap between the logical states $|1,\Uparrow\rangle$ and $|1,\Downarrow\rangle$. The dependence of the qubit coupling to such environment on the device geometry and field orientation might be correlated to that of the Rabi frequency $f_R^Z$. If this were the case, the above analysis should be complemented with that of the qubit-environment coupling, and $f_{R}^{XZ}$ should be replaced by a novel and more comprehensive figure of merit, able of identifying optimal operation points \cite{Wang2021a}.  

\section{Conclusion}

In conclusion, we have investigated the possibility of generating hole-spin qubits in a downscaled Si pMOSFET, derived from 22nm Fully-Depleted Silicon-on-Insulator CMOS foundry technology. We find that these devices indeed allow the formation of well-defined quantum dots, for a realistic set of values of the applied (sub-threshold) gate voltages. Besides, the application of time-dependent voltages to the top gate allows one to perform qubit rotations around two orthogonal axes, with Rabi frequencies of the order of $10^2\,$MHz (at $1\,$T). Despite the fact that the two rotations display complementary dependencies on the magnetic field orientation, good trade-offs have been identified, demonstrating the possibility of efficiently implementing the two qubit rotations within the same geometry. All the relevant quantities display a strong and non-monotonic dependence on the relative strength of the spatial confinement along the in-plane directions (length and width of the Si channel). In particular, for a channel width that is slightly larger than its length, a clear transition is observed in the ground-state properties, and is accompanied by a strong increase in the values of the maximal Rabi frequencies, which are obtained in the same range of magnetic field orientations.

\section*{Acknowledgements}

The authors acknowledge financial support from the European Commission through the project IQubits (Call:   H2020–FETOPEN–2018–2019–2020–01, ProjectID: 829005).

\appendix

\section*{Appendix A: TCAD simulations\label{app:tcad}}

The confining potential $U(\boldsymbol{r})$, has been computed with the TCAD software Ginestra\textsuperscript \textregistered, a novel multiphysics and multiscale Material-Device Simulation Platform, which establishes a functional link between material properties and device electrical performances. Ginestra\textsuperscript \textregistered relies on a kinetic Monte-Carlo engine and a solid physical description of the most relevant charge-transport (drift-diffusion, thermoionic emission at the interface between two media and generation/recombination rates) and material modifications mechanisms (degradation, polarization, phase change, and more) occurring during device operation. 
Accounting for the material properties, doping profile, and electrical parameters, the confining potential has been computed in the whole device and then extracted on the region of interest relative to the Si channel under the gate stack (see Fig.~\ref{fig:3DpotVG800}). The software computes the electrostatic potential by solving the Poisson and charge continuity equations, accounting for local electric fields, temperature and material parameters. 
One major advantage in using Ginestra\textsuperscript \textregistered is related to the capability to solve electrostatic equations even in a very low-temperature regime, reaching convergence in numerical calculations down to 1K and below, which is the range of temperatures involved in this study.
Another important feature of the Ginestra\textsuperscript \textregistered software, lies in the capability of accounting for a discrete description of defects which may be present in the gate stack or in other insulating regions of the device and represent an important contribution to electrostatics and transport modification (trap-assisted-tunneling). As a matter of fact, noise, variability and reliability are generated by the presence of unwanted and unavoidable defects, which may influence the electrical properties of the device, including the quantum confining potential.

\setcounter{equation}{0}
\renewcommand{\theequation}{B\arabic{equation}}
\section*{Appendix B: Multiband calculation of the hole states\label{app:kdotp}}

The hole states are obtained by diagonalizing a 6 band L\"uttinger-Kohn envelope function Hamiltonian $H_{\boldsymbol{k} \cdot \boldsymbol{p}}$, which accounts for the coupling between the heavy-hole ($J=|M|=3/2$), light-hole ($J=3/2$, $|M|=1/2$), and split-off ($J=|M|=1/2$) bands at the $\Gamma$ point. In the basis 
$\left\{\left|\tfrac{3}{2}, \tfrac{3}{2}\right\rangle, 
        \left|\tfrac{3}{2}, \tfrac{1}{2}\right\rangle, 
        \left|\tfrac{3}{2},-\tfrac{1}{2}\right\rangle, 
        \left|\tfrac{3}{2},-\tfrac{3}{2}\right\rangle, 
        \left|\tfrac{1}{2}, \tfrac{1}{2}\right\rangle, 
        \left|\tfrac{1}{2},-\tfrac{1}{2}\right\rangle \right\}$, 
the Hamiltonian reads \cite{kpbook}:
\begin{widetext}
\begin{eqnarray}
    H_{\boldsymbol{k} \cdot \boldsymbol{p}} = 
    \left(
    \begin{array}{cccccc}
              P+Q &             -S &             R &             0 &    -S\tfrac{1}{\sqrt{2}} & R \sqrt{2}     \\
             -S^* &            P-Q &             0 &             R &    -Q \sqrt{2} & S \tfrac{\sqrt{3}}{\sqrt{2}}   \\
              R^* &              0 &           P-Q &             S & S^* \tfrac{\sqrt{3}}{\sqrt{2}}  & Q\sqrt{2} \\
                0 &            R^* &           S^* & P+Q           &  -R^* \sqrt{2} & -S^*\tfrac{1}{\sqrt{2}}  \\
    -S^*\tfrac{1}{\sqrt{2}} &  -Q^* \sqrt{2} & S \tfrac{\sqrt{3}}{\sqrt{2}} &    -R \sqrt{2} &  P+\Delta_{SO} &              0 \\
     R^* \sqrt{2} & S^* \tfrac{\sqrt{3}}{\sqrt{2}}  &  Q^*\sqrt{2} &   -S\tfrac{1}{\sqrt{2}} &              0 &  P+\Delta_{SO} \\
    \end{array}
    \right),
   \label{k.p Hamiltonian}
\end{eqnarray}
\end{widetext}
where $\Delta_{SO}=44\,$meV is the spin-orbit parameter in Si, and the operators appearing in the above matrix are given by
\begin{eqnarray}
   P &=& \frac{\hbar^2}{2m_0}\gamma_1 (k_x^2+k_y^2+k_z^2) \,, 
\end{eqnarray}
\begin{eqnarray}
   Q &=& \frac{\hbar^2}{2m_0}\gamma_2 (k_x^2+k_y^2-2k_z^2) \,, 
\end{eqnarray}
\begin{eqnarray}
   R &=& \frac{\hbar^2}{2m_0}\sqrt{3} [-\gamma_3(k_x^2-k_y^2)+2i\gamma_2 k_x k_y] \,,  
\end{eqnarray}
\begin{eqnarray}
   S &=& \frac{\hbar^2}{2m_0}2\sqrt{3} \gamma_3(k_x-ik_y)k_z \,. 
\end{eqnarray}
The presence of a confinement potential, resulting both from the band offset in the heterostructure and from the voltage applied to the top- and back-gates in the pMOSFET, is accounted for by adding the potential function $U(\boldsymbol{r})$ to all the diagonal elements in $H_{\boldsymbol{k} \cdot \boldsymbol{p}}$. 

The presence of a static magnetic field is accounted for by the Hamiltonian term $H_B$. This includes the Zeeman-Bloch contribution, given by
\begin{widetext}
\begin{eqnarray}
    H_{B,Z} = 
    \left(
    \begin{array}{cccccc}
        3\kappa B_z & \sqrt{3}\kappa B_\phi^* & 0 & 0 & \sqrt{\tfrac{3}{2}}(\kappa+1) B_\phi^* & 0 \\
        \sqrt{3}\kappa B_\phi & \kappa B_z & 2\kappa B_\phi^* & 0 & -\sqrt{2}(\kappa+1) B_z & \tfrac{1}{\sqrt{2}}(\kappa+1) B_\phi^* \\
        0 & 2\kappa B_\phi & -\kappa B_z & \sqrt{3}\kappa B_\phi^* & -\tfrac{1}{\sqrt{2}}(\kappa+1) B_\phi & -\sqrt{2}(\kappa+1) B_z \\
        0 & 0 & \sqrt{3}\kappa B_\phi & -3\kappa B_z & 0 & -\sqrt{\tfrac{3}{2}}(\kappa+1) B_\phi \\
        \sqrt{\tfrac{3}{2}}(\kappa+1) B_\phi & -\sqrt{2}(\kappa+1) B_z & -\tfrac{1}{\sqrt{2}}(\kappa+1) B_\phi^* & 0 & (2\kappa+1) B_z & (2\kappa+1) B_\phi^* \\
        0 & \tfrac{1}{\sqrt{2}}(\kappa+1) B_\phi & -\sqrt{2}(\kappa+1) B_z & -\sqrt{\tfrac{3}{2}}(\kappa+1) B_\phi^* & (2\kappa+1) B_\phi & -(2\kappa+1) B_z \\
    \end{array}
    \right),
   \label{k.p Hamiltonian}
\end{eqnarray}
\end{widetext}
where $B_\phi = B_x+iB_y$.
Besides, we include the paramagnetic ($H_{B,P}$) and the diamagnetic ($H_{B,D}$) contributions. These come from the substitution $k_\alpha \rightarrow k_\alpha + \tfrac{e}{\hbar} A_\alpha$, where $\boldsymbol{A} = - \tfrac{1}{2} \boldsymbol{r} \times \boldsymbol{B}$ is the vector potential associated with the external magnetic field. The paramagnetic and diamagnetic Hamiltonians collect, respectively, the terms of order 1 and 2 in the magnetic field $\mathbf{B}$.

After the substitution of $\boldsymbol{k} \rightarrow -i\boldsymbol{\nabla}$ in $H_{\boldsymbol{k} \cdot \boldsymbol{p}}$ and $H_B$, one obtains a differential Schr\"odinger equation, which we diagonalize in the basis $|\boldsymbol{n},J,M\rangle$, where $|\boldsymbol{n}\rangle = |n_x,n_y,n_z\rangle$ are the eigenstates of a three-dimensional harmonic oscillator with mass $m_0/\gamma_1$ and optimized values of the angular frequencies $\omega_\alpha$.  

\setcounter{equation}{0}
\renewcommand{\theequation}{C\arabic{equation}}
\section*{Appendix C: Rabi frequencies\label{app:rabi}}

For the sake of the following discussion, it is convenient to expand a hole eigenstate as:
\begin{equation}\label{eq:bandexp}
    |m,\xi\rangle = \sum_{b} |\psi_{m,\xi}^{b}\rangle\otimes |b\rangle \,,
\end{equation}
where $|\psi_{m,\xi}^{b}\rangle$ are the unnormalized and (in general) non-mutually orthogonal envelope functions of the hole eigenstate corresponding to the band $b=(J,M)$. 
At zero magnetic field, the two eigenstates $|m,\xi\rangle$ that belong to the $m$-th doublet are one the time-reversal conjugate of the other: 
$ |m,\Uparrow\rangle = \Theta |m,\Downarrow\rangle$ and $ |m,\Downarrow\rangle = - \Theta |m,\Uparrow\rangle$. Here, $\Theta$ is the time-reversal operator, which acts on the eigenstate components as
\begin{equation}\label{fXcontr}
  \Theta |\psi_{m,\xi}^{b}\rangle \otimes |J,M\rangle = (-1)^{J+M} \big| (\psi_{m,\xi}^{b})^* \big> \otimes\big| J, -M \big>    \,.
\end{equation}
Since the potential operator is diagonal with respect to the band index, one can always express the overall Rabi frequency as the algebraic sum of six contributions, each coming from a particular band: 
\begin{equation}\label{fxrapp}
f_R^{X}=\frac{1}{h}\left|\sum_b \langle \psi_{1,\Uparrow}^{b} |\delta U|\psi_{1,\Downarrow}^{b} \rangle \right| \,.
\end{equation}
From the above Eqs. (\ref{eq:bandexp}-\ref{fXcontr}) it follows that $f_R^X$ vanishes identically in the zero-field limit, because the contributions $(J,M)$ and $(J,-M)$ cancel each other. 

Along the same lines, one can show that also $f_R^Z$ vanishes identically in the zero-field limit. In fact, from Eq. (\ref{fXcontr}) and from the expression 
\begin{equation}\label{fzrapp}
f_R^{Z}=\frac{1}{h}\left|\sum_b \left(\langle \psi_{1,\Uparrow}^{b} |\delta U|\psi_{1,\Uparrow}^{b} \rangle - \langle\psi_{1,\Downarrow}^{b} |\delta U|\psi_{1,\Downarrow}^{b} \rangle\right) \right| 
\end{equation}
it follows that the terms corresponding to $\xi=\Uparrow$ and $\xi=\Downarrow$, for each $(J,M)$, cancel each other.
 
Finally, the above Eqs. (\ref{fxrapp}-\ref{fzrapp}) allow one to estimate the direct contribution of each band to the Rabi frequencies. In particular, the direct contribution of the split-off band is given by the terms in that correspond to $b=(1/2,\pm 1/2)$. The indirect contribution of the split-off bands results instead from their effect on the mixing between the light- and heavy-hole bands in the ground states. 


\setcounter{equation}{0}
\renewcommand{\theequation}{D\arabic{equation}}
\section*{Appendix D: Larmor and Rabi frequencies in the $g$-matrix formalism\label{app:gmf}}

Within the $g$-matrix formalism, the effective qubit Hamiltonian reads:
\begin{align}
\mathcal{H} = \frac{1}{2} {\mu_B} \boldsymbol{\sigma} \cdot\hat{g} (V_G) \cdot \boldsymbol{B}\,,
\label{eq:htwol}
\end{align}
where $\boldsymbol{\sigma}$ is the vector formed by the Pauli matrices. The $g$ tensor depends on the confinement potential and thus (crucially) also on the top-gate voltage $V_G$. It can be derived from the expression of the hole eigenstates through the expression \cite{Venitucci2018a}:
\begin{align}
\hat{g}\!=\!\frac{2}{\mu_B}\!
\begin{pmatrix}
\text{Re}[\langle \Downarrow\! | M_x|\!\Uparrow\rangle] & \text{Re}[\langle \Downarrow \!| M_y|\!\Uparrow\rangle] & \text{Re}[\langle \Downarrow\! | M_z|\!\Uparrow\rangle] \\
\text{Im}[\langle \Downarrow\! | M_x|\!\Uparrow\rangle] &\text{Im}[\langle \Downarrow\! | M_y|\!\Uparrow\rangle] & \text{Im}[\langle \Downarrow\! | M_z|\!\Uparrow\rangle] \\
\langle \Uparrow\! | M_x|\!\Uparrow\rangle & \langle \Uparrow\! | M_y|\!\Uparrow\rangle &\langle \Uparrow\! | M_z|\!\Uparrow\rangle \\
\end{pmatrix}\,.
\nonumber
\end{align}
The components $M_\chi=\partial H_B /\partial B_\chi$ and the Pauli matrices are defined with respect to an arbitrary chosen reference frame and basis of hole states $\{|\!\Uparrow\rangle,|\!\Downarrow\rangle\}$ (belonging to the zero-field ground doublet). 
For small amplitudes of the oscillating gate potential ($ | \delta V_G | \ll | V_G |$), the first derivative of the $g$ matrix is numerically derived from the equation
\begin{align}
\hat{g}'(V_G) \approx \frac{1}{\delta V_G} [\hat{g}(V_G+\delta V_G) - \hat{g}(V_G)]\,.
\end{align}

With respect to the so-called \textit{principal magnetic axes} and for a suitable pair of qubit basis states, the $g$ matrix can be written in a diagonal form. Our calculations show that the principal magnetic axes approximately coincide with the crystallographic directions $[110]$, $[\overline{1}10]$, and $[001]$, which coincide here with the coordinate axes $x$, $y$, and $z$. The principal $g$ factors are derived by diagonalizing the symmetric Zeeman tensor $\hat{G}={}^t\hat{g}\cdot\hat{g}$, which is derived from the field-induced splitting for 6 different field orientations \cite{Venitucci2018a,Crippa2018a}. The eigenvectors of $\hat{G}$ identify the principal magnetic axes (at a given gate voltage), while the eigenvalues provide the moduli (but not the sign) of the principal $g$ factors \cite{Chibotaru2008a}. Here, for the calculations performed for $B=1\,$T and $V_{G}=-0.6,\,-0.8\,$V, we find that the principal magnetic axes deviate from the crystallographic ones $[110]$ ($x$), $[\overline{1}10]$ ($y$), and $[001]$ ($z$) by no more than $10^{-3}\,$rad, and can thus be identified with such axes.

The effective Hamiltonian can thus be expressed in terms of the $g$ factors $g_x$, $g_y$, and $g_z$ [Eq.~(\ref{eq:Hgdiag})], and the reference states $|\!\!\Uparrow\rangle$ and $|\!\!\Downarrow\rangle$ correspond to the qubit eigenstates for a weak $B$ field, oriented along the $z$ axis. The general expressions given in Eqs. (\ref{eq:freX}-\ref{eq:freZ}) can be simplified by assuming that the field lies along one of the coordinate planes. In particular, for $\phi=90^\circ$ ($yz$ plane), the Rabi frequency corresponding to rotations around the $X$ axis becomes:
\begin{align}
f_R^{X}(\theta)=&\frac{\mu_BBV_{ac}}{2h} \frac{|g_yg'_z-g_zg_y'||\sin\theta\cos\theta|}{\sqrt{g_y^2\sin^2\theta+g_z^2\cos^2\theta}}\,.
\label{eq:frxgendiag}
\end{align}
The derivative of the above $f_R^{X}$ with respect to $\theta$ vanishes for $\cos^2\theta=|g_x|/|g_x+g_z|$, hence the condition given in Eq.~(\ref{eq:thetamax}).

One can proceed along the same lines for the Rabi frequency $f_R^{Z}$. For $\phi=90^\circ$ ($yz$ plane), its expression becomes:
\begin{align}
f_R^{Z}(\theta)=&\frac{\mu_BBV_{ac}}{2h} \frac{|g_yg_y'\sin^2\theta+g_zg_z'\cos^2\theta|}{\sqrt{g_x^2\sin^2\theta+g_z^2\cos^2\theta}}\,.
\label{eq:frxgendiag}
\end{align}
The angles corresponding to the maximum and to the zero are readily found from Eq.~(\ref{eq:frxgendiag}). The equations that describe the dependence of $f_R^{X}$ and $f_R^{Z}$ on the field orientation in the $xz$ plane are derived from the above ones by replacing $g_y$ and $g_y'$ with $g_x$ and $g_x'$, respectively.


%

\end{document}